# Shifts in Rossby wave regimes via the change of group velocity with zonal jets

**Zhixiang Li[1,2], Jianhua Lu[1], Yimin Liu[2], and Tim Woollings[3]**

[1]School of Atmospheric Sciences, Sun Yat-sen University and Southern Marine Science and Engineering Guangdong Laboratory (Zhuhai), Zhuhai, China

[2]Key Laboratory of Earth System Numerical Modeling and Application, Institute of Atmospheric Physics, Chinese Academy of Sciences, Beijing, China

[3]Atmospheric, Oceanic and Planetary Physics, University of Oxford, Oxford, United Kingdom

Corresponding author: J. H. Lu (lvjianhua@mail.sysu.edu.cn), Y. M. Liu (lym@lasg.iap.ac.cn)

**Abstract**

Evidences show it is questionable to use the trends in Rossby wave phase speed ($C_{px}$) to interpret the relation between the changes of extreme events and westerly jets. We propose, by using 70 years (1954–2023) of ERA5 reanalysis, that the trends in extremes are associated with the shifts of Rossby wave regimes, via the westerly-induced changes in Rossby wave group velocity ($C_{gx}$), which are two to three times larger than those in $C_{px}$. We find significant deceleration in the subtropics and polar regions, alongside a salient midlatitude acceleration, of Rossby wave $C_{gx}$, especially over the East Asia–Pacific–North America and North Atlantic–Europe sectors. These changes induce robust shifts in spatiotemporal regimes of Rossby waves, from the transient synoptic-scale regime toward the quasi-stationary planetary-scale wave regime where $C_{gx}$ decelerates, and vice versa. The CESM2 Large Ensemble confirms that the observed pattern over the Eurasia–Pacific–North America sector is a forced response to external forcing, while uncertainties remain over the polar regions and the North Atlantic. We suggest that westerly trends may reshape the regimes of extratropical Rossby wave activity and extreme weather events through the changes in group velocity rather than phase speed of Rossby waves.

## 1 Introduction

Over the past decades, many extreme weather and climate events have become more frequent, and in some cases more persistent, with each type of extreme such as heat waves, cold spells, drought, flooding, and wildfires, having its own unique spatiotemporal pattern of the changes in frequency and persistence (e.g., Dunn et al., 2020; Perkins-Kirkpatrick & Lewis, 2020; White et al., 2022; Blackport et al., 2024; Graversen et al., 2025). While some of the regional changes in extreme events may be related to internal decadal (or multidecadal) variability, it is widely believed that changes in extreme events are mainly associated with human-induced global climate change (Seneviratne et al., 2021).

For the extreme events in the extratropics - due to the essential role of ubiquitous Rossby waves - it was proposed that the stronger and more persistent extreme events are caused by slower progression of Rossby waves due to weakened zonal wind and increased wave amplitude (Francis and Vavrus, 2012). However, this may be an oversimplified argument (Wallace et al., 2014), and more analyses suggest the relation between the change of the Rossby wave phase speed ($C_p$) and zonal wind is not robust (Barnes, 2013; Riboldi et al., 2020). Recently, Wu and Lu (2023) and Li et al. (2025) further pointed out that the weak trend of (transient) Rossby waves' phase speed is due to the opposite effects on $C_p$ of trends in zonal wind and associated meridional gradient of potential vorticity ($PV_y$).

Given the robust changes in background zonal winds under global warming (Simmons, 2022; Woollings et al., 2023; Ossó et al., 2025), a natural inference of Wu and Lu (2023) and Li et al. (2025) is that there could be salient trends in group velocity ($C_g$) of transient Rossby waves. This is because the zonal wind and associated $PV_y$ terms act in the same sense for the group velocity for most of Rossby waves, while they act oppositely for the phase velocity, as clearly shown in the following equations for a quasi-geostrophic baroclinic Rossby wave with its zonal, meiridional, and vertical wave numbers being *k, l, m*, respectively:

$$C_{px} = \frac{\omega}{k} = \bar{u} - \frac{PV_y}{k^2+l^2+\frac{f^2}{N^2}(\frac{1}{4H^2}+m^2)} \quad (1)$$

$$C_{gx} = \frac{\partial\omega}{\partial k} = \bar{u} + \frac{PV_y\left[k^2-l^2-\frac{f^2}{N^2}(\frac{1}{4H^2}+m^2)\right]}{\left[k^2+l^2+\frac{f^2}{N^2}(\frac{1}{4H^2}+m^2)\right]^2} \quad (2)$$

$$PV_y = \beta - \bar{u}_{yy} - \frac{\partial}{\partial p}\left(\frac{p^2f^2}{N^2H^2}\frac{\partial\bar{u}}{\partial p}\right) \quad (3)$$

in which $\bar{u}$, $f$, $\beta$, $N$, and $H$ denote background zonal wind, Coriolis parameter and its meridional gradient, Brunt-Väisälä frequency, and atmospheric scale height, respectively (e.g., Phillips, 1990; Wu and Lu, 2023). The group velocity, i.e., the propagation velocity of wave packets or the velocity of energy dispersion, has long been recognized as one of the most fundamental properties of atmospheric waves (Rossby, 1945; Yeh, 1949; Phillips, 1990).

Extratropical large-scale Rossby waves involve different wave regimes in terms of their spatiotemporal scales and dynamical origins, including transient waves (2–9 days) and quasi-stationary waves (above 10 days) in time scale, and planetary waves (zonal 1–3 waves) and synoptic waves (zonal 4–15 waves) in spatial scale. The dynamics of transient synoptic Rossby waves, including their three-dimensional propagation, instabilities, and nonlinear interaction with the background flow, and their role in maintaining atmospheric general circulation, has been well developed since the era of Rossby and Charney. The underlying dynamics of (quasi-) stationary waves and planetary-scale waves are much more diverse, involving the teleconnections caused by thermal and topographic forcings, nonlinear interaction of transient eddies, etc., but they are key to understanding the trend of extreme events.

Rossby waves are usually represented in the global (i.e., domainwise) spectral frequency ($\omega$) – wave number ($K$) space, and accordingly the local weather condition can be represented by the Fourier integration/summation of all of wave modes at that location (Riboldi et al., 2020; Wicker et al., 2024). In this way, the persistence or the transition of local/regional weather regimes, or the temporal and regional variations in the propagation of Rossby waves, cannot be easily and directly shown in the global $\omega$–$K$ space representation of Rossby waves. However, a spectral representation of Rossby waves in localized $\omega$–$K$ space (Fragkoulidis & Wirth, 2020; Li et al., 2025) may provide local wave frequency, wave numbers (and the phase and group velocities of the waves), and hence can overcome the weakness in global spectral representation of Rossby waves. Indeed, efforts have been made to diagnose the climatology of local $C_{gx}$ from the zonal movement speed of Rossby wave packets based on lag-correlation maps (Chang & Yu, 1999, CY99 hereafter) and the analytic signal of Rossby wave packet envelopes (Fragkoulidis & Wirth, 2020, FW20 hereafter).

**Here we propose that the debate on the slower progression of Rossby waves or the change in waviness of atmospheric circulation can be envisaged instead from an alternative perspective, i.e., the shifts of Rossby wave spatiotemporal-scale regimes associated with the**

**changes in Rossby wave group velocity which is in turn closely related to the changes in zonal jets**. By developing a novel method to estimate $C_{gx}$ based on $\partial\omega/\partial k$ and comparing it with CY99 and FW20, we first quantify how the spatial pattern of $C_{gx}$ responds to zonal wind trends over the period from 1954 to 2023. The paper is structured as follows: Section 2 presents the data and methodology, Section 3.1 reveals the differences in the responses of $C_{px}$ and $C_{gx}$ to local zonal wind trend, Section 3.2 describes the spatial pattern of trend in group velocity, Section 3.3 describes the relation between $C_{gx}$ trends and shift in Rossby wave regimes. The summary and discussion are presented in Section 4.

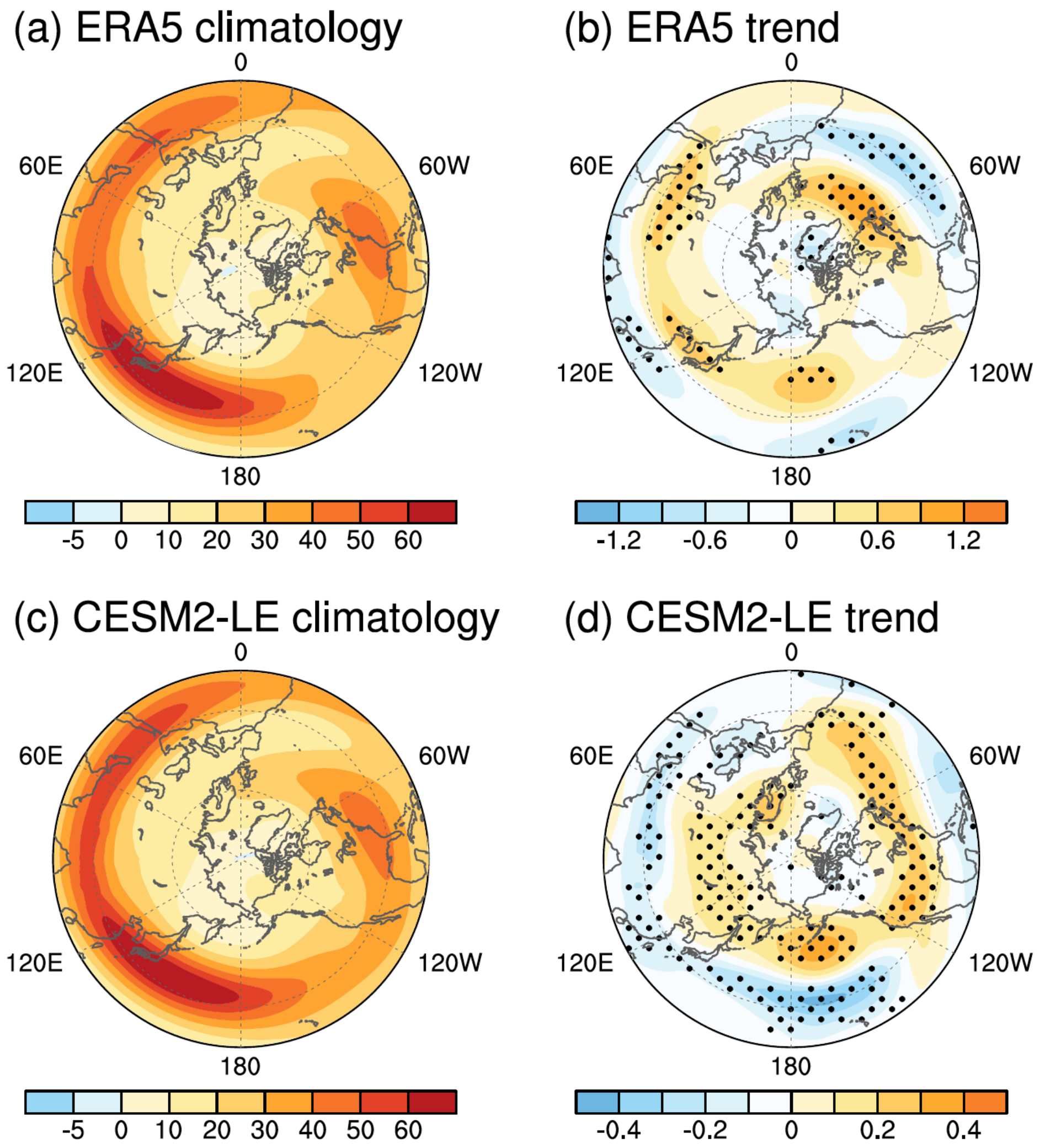

**Figure 1. Observed and modeled climatology and trends of U200 during boreal winter from 1954 to 2023.** (a–b) Climatology (a; units: m/s) and trend (b; units: m/s/decade) of DJF mean U200 based on ERA5. Stippling indicates trend is statistically significant at the 95% confidence level. (c–d) Climatology (c; units: m/s) and trend (d; units: m/s/decade) of DJF mean U200 based on the CESM2-LE. Stippling indicates the regions where at least 80% members agree on the sign of the trend.

## 2 Data and Methods

### 2.1. Data

In this study, we use zonal wind ($u$) and meridional wind ($v$) at 200 and 500 hPa from the ERA5 reanalysis data (Hersbach et al., 2020) over the period 1954–2023 (70 years) with a 6–hourly temporal resolution (daily at 0000, 0600, 1200 and 1800 UTC) and a horizontal resolution of 2.5°×2.5°. Additionally, 10 realizations from the CESM2 Large Ensemble (CESM2-LE, Rodgers et al., 2021) that output 6-hourly $v$ and monthly-mean $u$, which comprise historical (1954–2014) and future (2015–2023) simulations, are used in this study. The historical simulations are forced by observed radiative forcing, while future simulations follow the SSP3-7.0 emissions scenario. The use of a large ensemble enables the separation of the forced response under global warming (ensemble mean) from the internal climate variability (spread of ensemble members) (Huang et al., 2020; Kang et al., 2025). All data in the CESM2-LE are interpolated to the same horizontal resolution and pressure levels as ERA5 to facilitate the comparison. We focus on boreal winter, defined as the December–February (DJF) mean. Accordingly, the long-term trends in this study are calculated using 69 winter seasons over the 70-year data, while results for the satellite era (1979–2023) are based on 44 winters.

### 2.2. Local envelopes, angular frequency, angular wavenumber and $C_{px}$

Spatiotemporal local wave parameters (inluding the envelopes of wave packets, frequency, and wavenumber) of extratropical Rossby waves are obtained by applying the Hilbert transform in the zonal direction on 6-hourly $v$ (Fragkoulidis & Wirth, 2020; Li et al., 2025). Specifically, we construct the complex signal $V$ by defining it as $V=v+i\hat{v}$, where $\hat{v}$ represents the convolution-based Hilbert transform of $v$. The local phase $\Phi$ and local envelopes $E$ are then derived by computing the polar angle and the polar radius, respectively, of the polar form of the complex signal $V$, i.e., $V=Ee^{i\Phi}$, where $E=\sqrt{v^2+\hat{v}^2}$, and $\Phi=\tan^{-1}(\hat{v}/v)$. Therefore, the local angular frequency $\omega$ and zonal angular wavenumber $k$ can be calculated from

$$\omega=-\frac{\partial\Phi}{\partial t},k=\frac{\partial\Phi}{\partial x}=\frac{1}{a\cos\varphi}\frac{\partial\Phi}{\partial\lambda} \quad (4)$$

where notations are standard. The zonal phase speed is then given by

$$C_{px}=\frac{\omega}{k} \quad (5)$$

Following previous studies, the $C_{px}$ and $C_{gx}$ in this paper are diagnosed based on the synoptic-scale $v$ (zonal wavenumbers 4–15; CY99; FW20). In addition, wave parameters for zonal wavenumbers 1–15 of $v$ are also diagnosed to distinguish the spatiotemporal-scale regimes of extratropical Rossby waves. Full technical details can also be found in Li et al. (2025).

### 2.3. Three different $C_{gx}$ metrics

The climatology of $C_{gx}$ and its long-term linear trends are calculated using three different metrics:

1. The CY99 method: This metric detects the seasonal mean $C_{gx}$ by calculating the zonal propagation speed of $E$ derived from one-point lag-correlation maps, based on the local maxima in the correlation coefficients from time lag of hours -12 to hours +12.
2. The FW20 method: This metric detects the local $C_{gx}$ by calculating the zonal propagation speed of wave packet segments based on the Hilbert transform. As shown in FW20, the procedure is analogous to that used for detecting $C_{px}$ (section 2.2), except that the target variable is the 4000-km high-pass-filtered local envelope $E$, and the diagnosis is performed on the $E$ segments. The $E$ segments are defined to occur when $E$ exceeds $E_0$ (12.5 m/s in this study) for at least 20° in longitude, and the corresponding segments at preceding and succeeding time steps are tracked to eliminate spurious $C_{gx}$ caused by growth or decay of $E$. The mean amplitude of each segment is then subtracted (i.e., $E' = E - \bar{E}$). Finally, the local phase ($\Phi_{E'}$), angular frequency ($\omega_{E'}$) and zonal angular wavenumber ($k_{E'}$) of the $E'$ are calculated following the procedure described in section 2.2, and $C_{gx}$ is obtained as $\omega_{E'}/k_{E'}$. However, as noted by FW20, this method still has limitations when two or more local maxima exist within the wave packet segment. Substantial missing values are introduced by the threshold $E_0$. Therefore, in this study, the FW20 method is used primaily as a reference for the other two methods.
3. $\partial\omega/\partial k$: We develope a new method for estimating the local $C_{gx}$ from the original definition of group velocity, $C_{gx} = \partial\omega/\partial k$. To this end, we firstly set $\omega$ to be positive definite and the corresponding $k$ to be negative when a wave propagates westward. An exemplary $\omega$–$k$ spectrum for 1980/81 DJF at 30°N is shown in Fig. S2a. Given the distinct linear characteristics of the $\omega$–$k$ spectrum, we assume $\omega$ can be linearly expressed as a function of $k$ at each grid point and time step, and we compute the linear regression coefficient between variations in local frequency and in local zonal wavenumber within a narrow range of space and time, i.e.,

$$\omega \approx \omega_0 + \frac{\partial \omega}{\partial k}(k - k_0) \qquad (6)$$

where $\omega_0, k_0$ are the local angular frequency and local zonal angular wavenumber at which the $C_{gx}$ is estimated, $\partial\omega/\partial k$ represents the local $C_{gx}$. The $C_{gx}$ is calculated using the linear regression via least-squares from $\omega$–$k$ pairs that satisfy: (a) within a spatial range of 7.5° ×7.5°; (b) within a temporal range of ±24 hours; (c) satisfy $\left|\frac{k-k_0}{k_0}\right| < 0.5$. Only the linear regression coefficients that are statistically significant at the 95% confidence interval level are saved. Fig. S2b shows an instantaneous example of $\omega$–$k$ pairs centered at the grid point of 15°E, 30°N at 00UTC 15 January 1981, with a significant linear regression coefficient (i.e., $C_{gx}$).

In the end, for $C_{px}$ and $C_{gx}$ based on the FW20 method, grid points where the absolute values exceed 50 m/s and 100 m/s are masked, respectively, and the median is used to represent seasonal mean (Fragkoulidis & Wirth, 2020). For all metrics, the seasonal mean is not calculated when undefined values are more than 80% during the season, and the long-term linear trend are only computed when seasonal mean is defined for more than 48 years during 1954–2023. The statistical significance is assessed based on a two-tailed Student's *t*-test. All results are robust to small variations in above mentioned parameters.

## 2.4. Probability density distribution of spatiotemporal-scale regimes

To investigate the climatological distribution and shifts of the spatiotemporal-scale regimes of extratropical Rossby waves, the local wave periods (i.e., $2\pi/|\omega|$; Li et al., 2025) derived from zonal wavenumers 1–15 $v$ are categorized into three different timescales at each grid point: short (2–9 days), intermediate (10–30 days), and long (>30 days), following the classification of Blackmon et al. (1984). For clarity, waves with periods of 2–9 days are also referred to as transient wave regimes, while those with periods of 10–30 days and >30 days are referred to as quasi-stationary wave regimes. Similarly, the spatiotemporal local zonal wavenumbers are categorized into planetary scale (zonal wavenumer 1–3) and synoptic scale (zonal wavenumer 4–15) regimes.

## 3 Results

### 3.1. Trends in $C_{gx}$ and $C_{px}$: Notable Difference in Magnitude

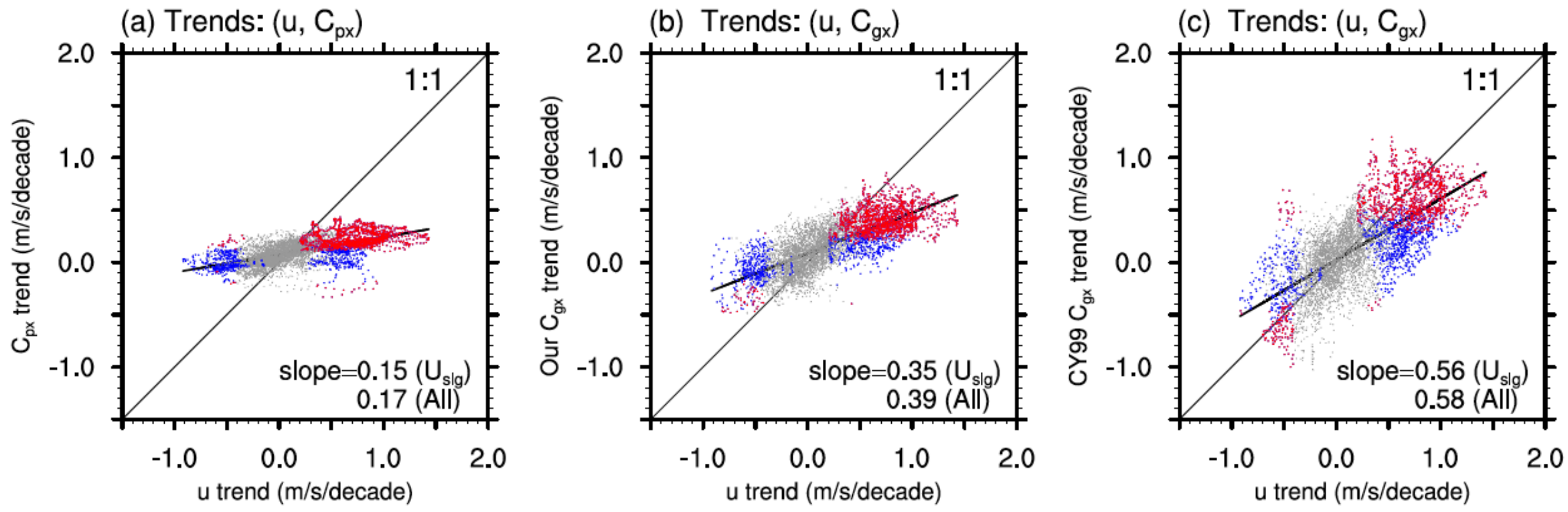

**Figure 2. The relationship between local trends in $C_{px}$ (a) or $C_{gx}$ (b–c) and local $u$ trends during DJF over 1954–2023 at 200 hPa in ERA5, based on all grid points in the extratropics (between 30°–80°).** The $C_{gx}$ trends are obtained from our new method (b) and CY99 method (c). The grid points with significant $u$ trends are marked by larger blue dots, while those exhibiting both significant $u$ trends and $C_{px}$ (a) or $C_{gx}$ (b–c) trends are further highlighted in red. The thick black line in each panel shows the linear regression line calculated from all dots, with its slope shown in the lower right corner. For comparison, the slopes computed from dots with significant u trends are also given.

Statistically significant zonal wind trends are emerging in both ERA5 and CESM2-LE, characterized by subtropical deceleration and midlatitude acceleration (Fig. 1). Figure 2 shows the different responses of local $C_{px}$ trend and local $C_{gx}$ trend (estimated from $\partial\omega/\partial k$ and the CY99 method) to the local $u$ trend. Both the $C_{px}$ and $C_{gx}$ trends exhibit positive correlations with the $u$ trend. However, the slopes of the linear regression indicate that the $C_{gx}$ trends are approximately two to three times the $C_{px}$ trends, which is expected given the distinct effects of changes in $PV_y$ on $C_{px}$ and $C_{gx}$, as outlined in the Introduction. Consequently, under global warming, changes in the zonal wind will induce significant variations in the group velocity of Rossby waves. Their spatial patterns and associated impacts will be analyzed in the subsequent sections.

### 3.2. Spatial Patterns of $C_{gx}$ from Different Metrics: Climatology and Trend

The spatial patterns of the climatology and long-term linear trend of $C_{gx}$ and $u$ during DJF from ERA5 and CESM2-LE are shown in Figure 3. The new method, which estimates $C_{gx}$ from $\partial\omega/\partial k$, produces results similar to those obtained using the CY99 and FW20 methods. All three metrics indicate that the locations of the maximum $C_{gx}$ coincide with the core of the jet stream, which are recognized as the midlatitude Rossby waveguides (Hoskins & Ambrizzi, 1993). These waveguides serve as the primary domains for significant propagation of synoptic-scale Rossby wave packets during boreal winter (CY99; FW20). In addition, $C_{gx}$ is much weaker in the polar regions, both due to the weak mean flow and to the weak β-effect or PV gradient (Woollings et al., 2023b). Both the new method and the CY99 method capture negative $C_{gx}$ (westward energy propagation) over the Russian Far East (grey shading in Figs. 3a–b and d–e), a feature that has rarely been documented.

Despite some quantitative discrepancies, the three metrics reveal highly consistent spatial patterns in the $C_{gx}$ trends, which generally align with the signs of local $u$ trends. In ERA5 (Figs. 3g–i), significant positive trends are observed across the midlatitudes, most notably over the East Asia-North Pacific-northwestern North America sector and the North Atlantic, while significant negative trends appear in the subtropics and polar regions (particularly evident with the CY99 method; Fig. 3h). The spatial pattern of $C_{gx}$ slowing down in the subtropics but accelerating in the midlatitudes during DJF is generally consistent with the poleward shift of the zonal mean jet stream (Archer & Caldeira, 2008; Chen et al, 2020; Woollings et al., 2023a; Breul et al., 2025), though pronounced spatial inhomogeneity in the magnitudes of the $C_{gx}$ and $u$ trends exists across the Northern Hemisphere.

The forced responses of long-term trends in $C_{gx}$ and $u$ during 1954–2023, based on the ensemble mean of CESM2-LE, are shown in Figs. 3j–l. Consistent with ERA5, the forced response reproduces the clear slowdown of $C_{gx}$ and $u$ in the subtropics and acceleration in the mid-latitudes, particularly over the East Asia-North Pacific-northwestern North America sector, although the magnitudes are weaker and the northward shift is more pronounced. Notable differences appear over the North Atlantic, where the positive trends are located farther south than in ERA5 and extend toward the subtropics. This discrepancy may stem from the limited skill of climate models in simulating long-term trends over the North Atlantic (Blackport & Fyfe, 2022).

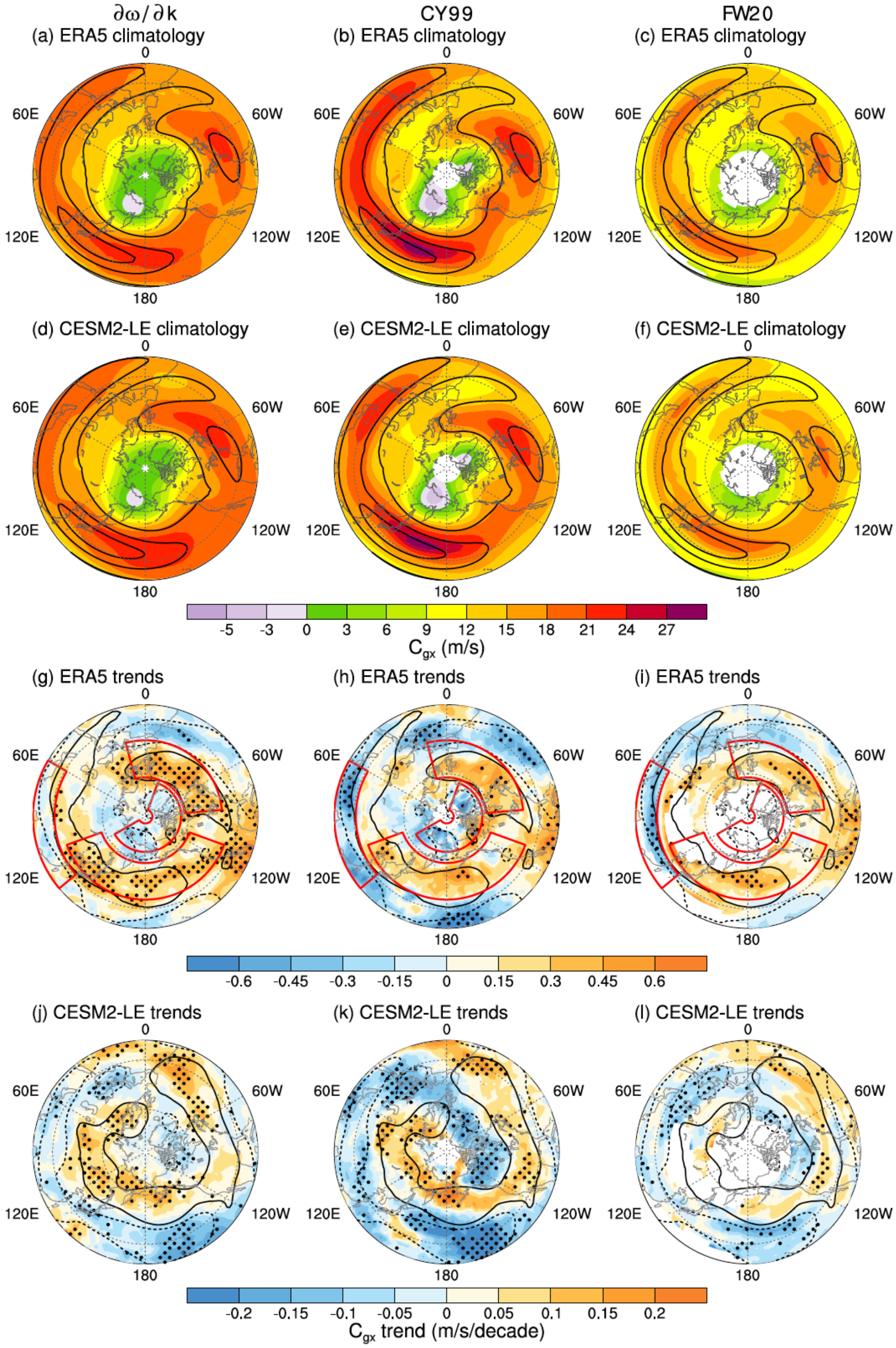

**Figure 3. Observed and modeled climatology and trends in $C_{gx}$ and $u$.** Climatology (a–c: ERA5; d–f: CESM2-LE; units: m s$^{-1}$) and trends (g–i: ERA5; j–l: CESM2-LE; units: m s$^{-1}$ (10 yr)$^{-1}$) for $C_{gx}$ during DJF over 1954–2023 at 200 hPa derived from our new method (left panel), CY99 method (middle panel) and FW20 method (right panel), respectively. Black contours indicate the corresponding climatology (a–f, only 20, 40, 60 m s$^{-1}$) and trends (g–i, only $\pm$0.25 m s$^{-1}$ (10 yr)$^{-1}$; j–l, only $\pm$0.1 m s$^{-1}$ (10 yr)$^{-1}$; negative contours dashed) of DJF mean $u$. Stippling indicates regions where trends in $C_{gx}$ are statistically significant above the 95% confidence level in ERA5 (g–i) or where at least 80% members of CESM2-LE agree on the sign (j–l). The color-bar interval of trends based on ERA5 is three times than that of CESM2-LE. The boundary circle is 20°N.

### 3.3. Connection between group velocity and spatiotemporal regime of Rossby waves

Strong group velocity and dispersion are unfavorable for the maintenance of quasi-stationary Rossby waves (Yeh, 1949; Luo et al., 2019a, b). The climatological group velocity therefore dictates the probability distribution of spatiotemporal scale of Rossby waves, while significant trends in group velocity would alter these distributions. To investigate this relationship, we first have a look at the climatology and trends of the probability density distribution (PDF) of wave periods during boreal winter in Figure 4. The climatological PDFs show that transient Rossby waves dominate in the midlatitudes with strong group velocity (Figs. 4a and d), whereas quasi-stationary Rossby waves prevail in high latitudes with weak group velocity (Figs. 4b–c and e–f). Moreover, the minimum PDF values for quasi-stationary waves in the midlatitudes (blue and purple shading in Figs. 4b–c and e–f) coincide with regions of maximum group velocity.

In the ERA5 reanalysis over 1954–2023, trends in the wave-period PDFs reveal a regime shift in terms of the changing balance between quasi-stationary and transient waves. The spatial pattern of trends in the PDF of the quasi-stationary wave regime (Figs. 4h–i combined) resembles that of trends in $C_{gx}$ (Figs. 3g–i), with an averaged pattern correlation coefficient (PCC) of -0.57. The negative PCC indicates that enhanced group velocity suppresses quasi-stationary waves, while weakened group velocity favors their occurrence. Specifically, the quasi-stationary wave regime shows significant decreases in the midlatitudes (Figs. S5b, c) but significant increases in the polar regions (Fig. S5a) and subtropics (Fig. S5d) (Figs. 4h and i), while the transient wave regime experiences the opposite changes (Fig. 4g).

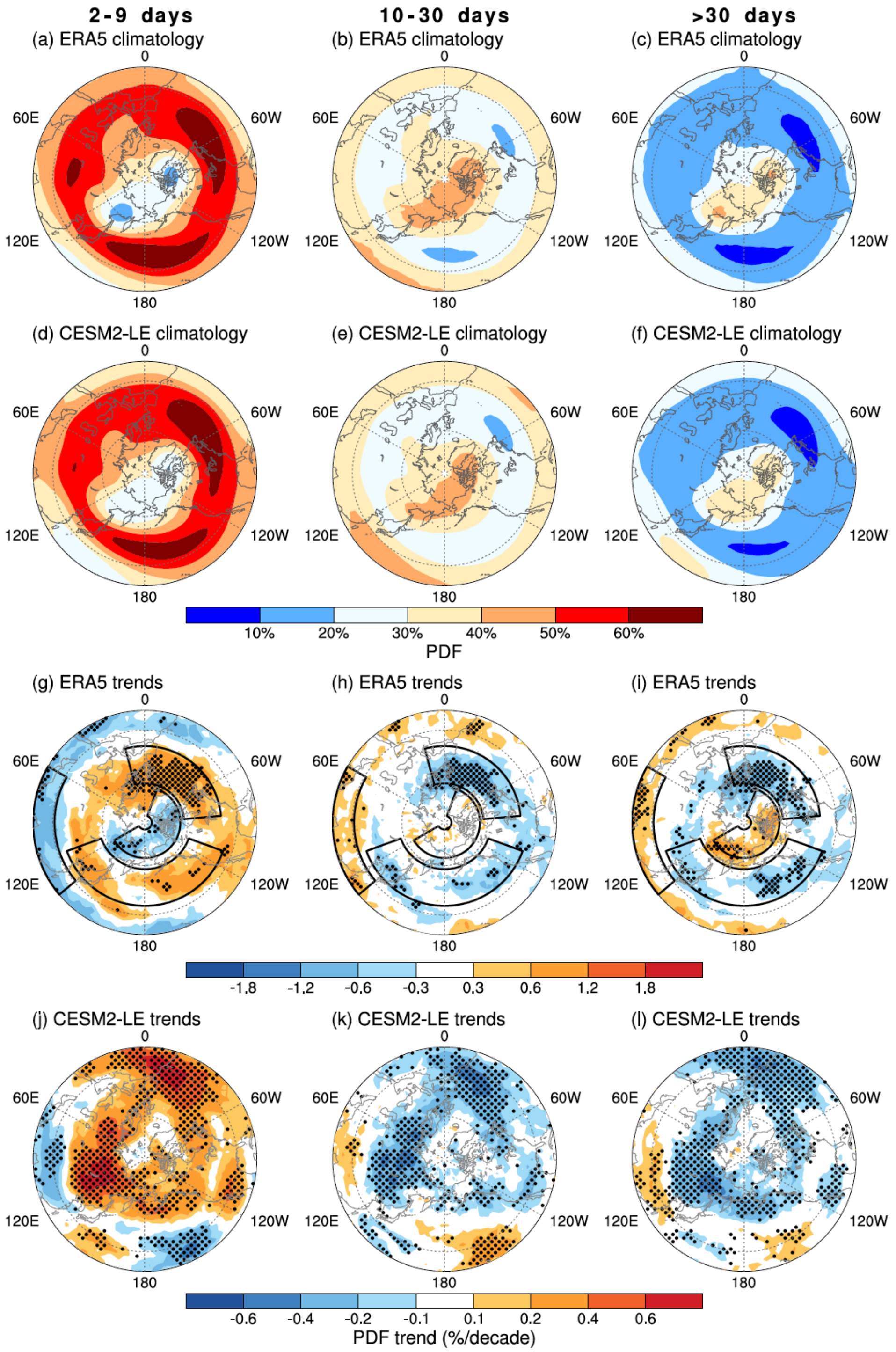

**Figure 4. Observed and modeled climatology and trends in the PDF of wave periods.** Climatology (a–c: ERA5; d–f: CESM2-LE) and trends (g–i: ERA5; j–l: CESM2-LE) for the probability density distribution of wave periods during DJF over 1954–2023 at 200 hPa. The three panels (from left to right) correspond to the bin of 2–9 days, 10–30 days, and >30 days, respectively. Stippling indicates regions where trends are statistically significant above the 95% confidence level in ERA5 (g–i) or where at least 80% members of CESM2-LE agree on the sign (j–l). The boundary circle is 20°N.

This shift in wave-period regimes is accompanied by a corresponding change in the PDF of zonal wavenumbers. In the subtropics and polar regions, where wave periods transition from transient to quasi-stationary waves, the wavenumber PDFs also shift from synoptic-scale toward planetary-scale waves (Figs. 5e–f; Figs. S5e and h). In contrast, in the midlatitudes, the decline in the quasi-stationary wave regime is associated with an increased proportion of synoptic-scale waves, particularly over East Asia and from the North Atlantic to northern Europe (Figs. 5e–f; Figs. S5f and g). However, we also find that although quasi-stationary waves are decreasing over the northeastern Pacific and northeastern North America, the associated wavenumber shift is not significant. This may be related to the fact that quasi-stationary waves can sometimes also be of synoptic scale (zonal wavenumbers 4–15; Röthlisberger et al., 2019; Fei & White, 2023; Li et al., 2025).

The shifts in the spatiotemporal regimes of Rossby waves are discernible from the time-lagged one-point correlation maps (Fig. S6), which show that, relative to the earlier 30-year period (1954/55–1983/84), a greater proportion of waves propagate toward northwestern North America and northern Europe in the recent 30 years (1993/94–2022/23). This reflects a change in the propagation patterns of Rossby waves. Such differences are also consistent with the northward shift of the storm tracks (Tamarin-Brodsky & Kaspi, 2017; Chemke & Yuval, 2026).

The forced responses simulated by the CESM2-LE largely reproduce the observed PDF trends, capturing a shift from the quasi-stationary planetary-scale wave regime toward the transient synoptic-scale wave regime in the midlatitudes, together with an opposite shift in the subtropics (mainly over South Asia and the Pacific jet exit region) (Figs. 4j–l; Figs. 5g–h). However, the regime shift in the polar regions revealed by ERA5 is not captured in the CESM2-LE ensemble mean.

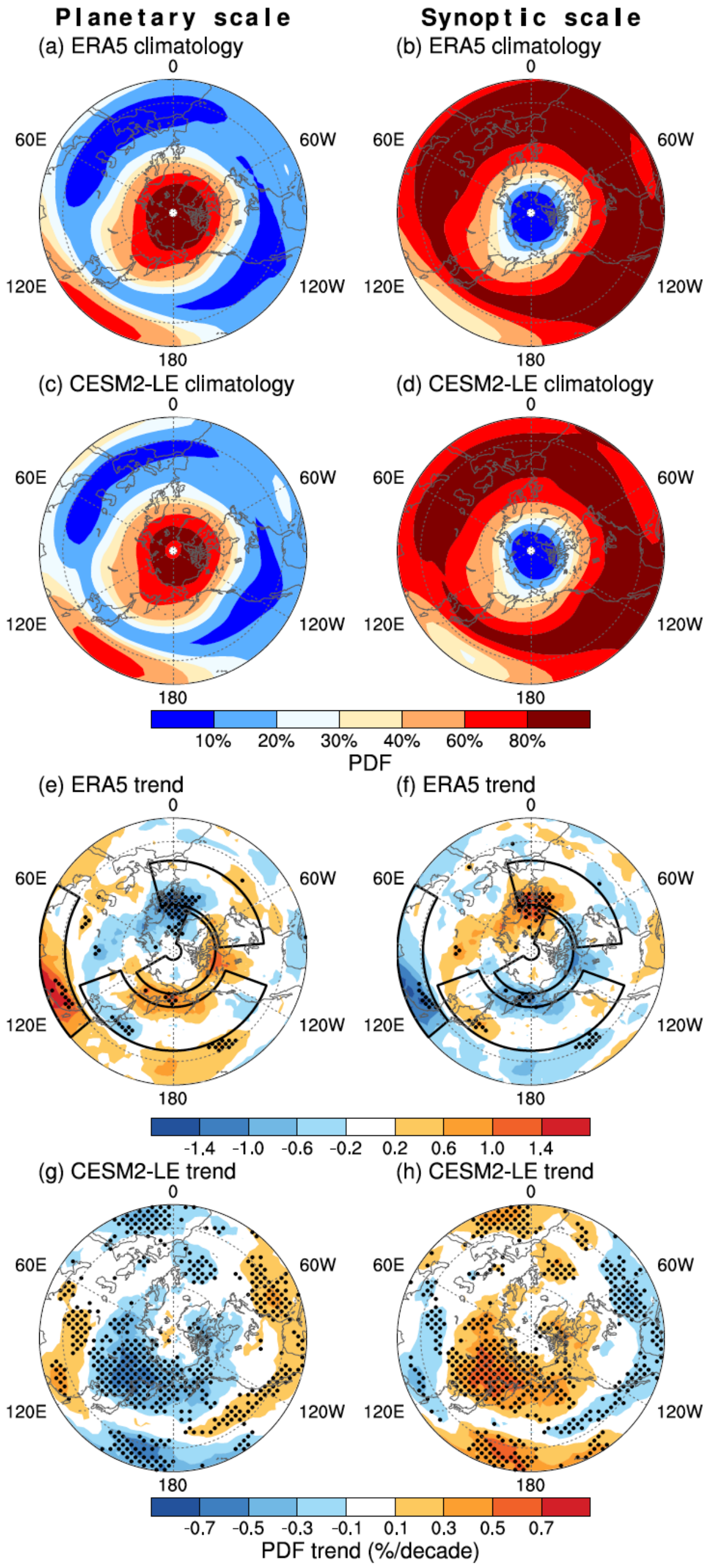

**Figure 5. Observed and modeled climatology and trends in the PDF of wavenumbers.** Climatology (a–b: ERA5; c–d: CESM2-LE) and trends (e–f: ERA5; g–h: CESM2-LE) over 1954–2023 for the probability density distribution of zonal wavenumbers during DJF at 200 hPa. The two panels (from left to right) correspond to the bin of planetary waves, and synoptic waves, respectively. Stippling indicates regions where trends are statistically significant above the 95% confidence level in ERA5 (e–f) or where at least 80% members of CESM2-LE agree on the sign (g–h). The boundary circle is 20°N.

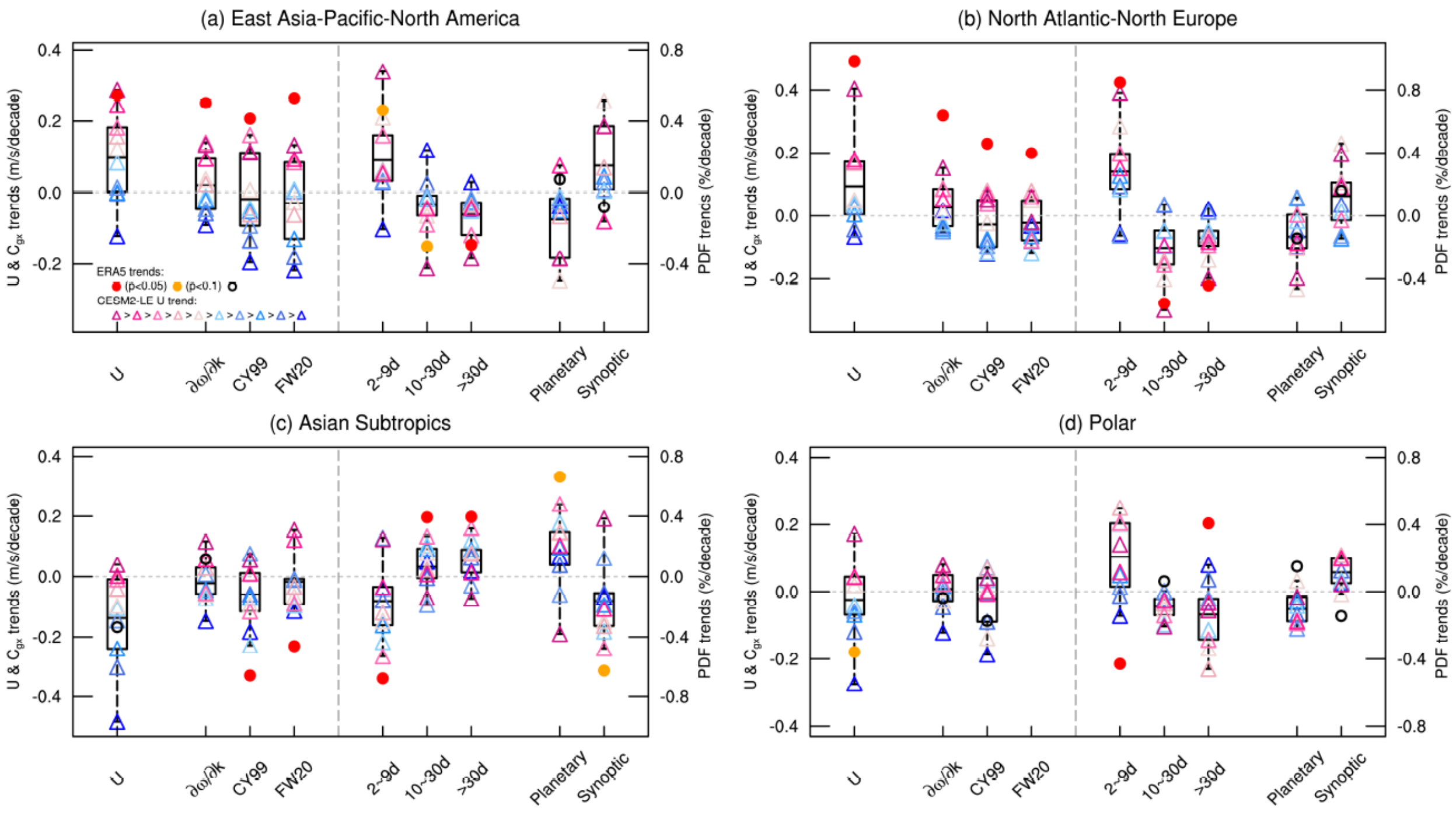

**Figure 6. Observed and modeled regional mean of the long-term trends over (a) East Asia–Pacific–northwestern North America, (b) North Atlantic–northern Europe, (c) the Asian subtropics and (d) the polar regions.** At the left side of the vertical gray dashed line in each panel are trends of zonal wind and the $C_{gx}$s estimated by three different methods (units: m s$^{-1}$ decade$^{-1}$), and at the right side are the PDFs of wave period and wavenumber (units: % decade$^{-1}$) All trends are calculated for DJF at 200 hPa over 1954–2023. Observed ERA5 trends are shown as circles and are filled in red (orange) if significant at 95% (90%) confidence level. Boxplots show the mean (central line) and the interquartile and full ranges of the CESM2-LE. Coloured triangles denote the ten individual ensemble members, ordered by the magnitude of the zonal wind trend in each panel. The color of each ensemble member remains the same across the other variables to highlight the relationship between zonal wind trends and the remaining quantities.

We summarize the main results in Figure 6, in which the left side of each panel shows the regional average of trends of zonal wind and $C_{gx}$, and the right side shows shifts of Rossby wave regimes in both the ERA5 reanalysis and CESM2-LE simulations. In the observations, both zonal wind and $C_{gx}$ accelerate over East Asia–Pacific–northwestern North America (Fig. 6a) and North Atlantic–northern Europe (Fig. 6b), accompanied by a significant shift of Rossby waves from quasi-stationary to transient wave regimes. The opposite results occur over the Asian subtropics (Fig. 6c) and the polar regions (Fig. 6d). The ten CESM2-LE ensemble members reproduce the relationship well, with stronger zonal wind trends linked to larger $C_{gx}$ changes and stronger shifts in the spatiotemporal scale of Rossby waves (shown by the coloured triangles in each panel). However, although the forced responses in CESM2-LE are consistent in sign with the observations in Figs. 6a–c, they are much weaker, being about one-third of the observed changes. In the polar regions the model produces the opposite shifts toward more transient synoptic waves (Fig. 6d), and the wave changes in ERA5 are outside of the range of the CESM2-LE. Relative to other variables, the observed $C_{gx}$ trends often fall outside the ensemble distribution, highlighting the need for improved model accuracy in simulating the response of Rossby wave dispersion.

The corresponding long-term trends in group velocity and the shifts in the spatiotemporal-scale regimes of extratropical Rossby waves at 500 hPa are shown in Figs. S7–S9 and are basically consistent with the results at 200 hPa. We also investigate the trends during the satellite era (1979–2023). Over this recent 45-year period, the deceleration of group velocity and zonal wind in the South Asian subtropics and polar regions (particularly east of the Ural Mountains), together with the northward shift over the Pacific-North America sector, becomes substantially stronger (Figs. S10g–i). Corresponding shifts in the spatiotemporal scale of Rossby waves, from transient synoptic-scale toward quasi-stationary planetary-scale waves where the flow decelerates and vice versa, are likewise evident in these regions (Figs. S11g–i and S12e–f). The increase in the quasi-stationary planetary-scale wave regime east of the Ural Mountains (60°–120°E) (Figs. S11h–i and Fig. S12e) aligns with the increasing blocking frequency trends observed in this region (Li et al., 2023). The forced response in CESM2-LE during the satellite era exhibits a similar signal over the North Pacific. In contrast, pronounced discrepancies with ERA5 appear over the South Asian subtropics, the North Atlantic, and the polar regions (Figs. S10j–l). Whether these differences arise from internal climate variability or from model biases warrants further investigation. Indeed, recent studies indicate that satellite-era trends over South Asia and the Pacific-North America sector are influenced by tropical Pacific SST trends (Kang et al., 2025), and that climate models may systematically underestimate Arctic climate change (Rantanen et al., 2022; Bao et al., 2026).

## 4 Summary and Discussion

Previous studies have focused on the $C_{px}$ trend to interpret regional weather change. However, by proposing a new method to detect local $C_{gx}$ from the calculation of $\partial\omega/\partial k$ and comparing it with two approaches based on Rossby wave packet speeds, we reveal that $C_{gx}$ is robustly changing two to three times faster than $C_{px}$, consistent with the theoretical expectation of distinct $PV_y$ impacts on $C_{px}$ and $C_{gx}$. It is the $C_{gx}$ trend determines the changes in the energy dispersion of Rossby waves, ultimately leading to shifts in the probability density between the two regimes of quasi-stationary and transient synoptic waves. In addition, the poleward shift in the maximum $C_{gx}$ over the North Pacific signifies the poleward shift of the storm track, which further determines the direction of energy and moisture transport (Tamarin-Brodsky & Kaspi, 2017; Wang et al., 2017; Li & Ding, 2024). Therefore, the changes in $C_{gx}$ provide a dynamic linkage between changes in zonal wind and changes in regional weather and climate under global warming. One motivation for studying changes in wave regime has been concerns over the potential for increased persistence associated with slower-moving waves. Our work shows that this is more likely to occur in subtropical as opposed to midlatitude regions.

Whether and how regional weather and climate in northern midlatitudes will be influenced by Arctic warming and tropical warming remains unclear (Screen et al., 2018; Cohen et al., 2020). Since the observed zonal wind trends resemble a poleward shift more than an equatorward shift, it is likely that tropical warming dominates. However, the uncertainties may be intensified by the ongoing debate on discrepancies between simulated El Niño-like and observed La Niña-like sea surface temperature trend pattern (Lee et al., 2022; Seager et al., 2022; Fu & Fedorov, 2023; Patterson & O'Reilly, 2025). The robust relationship between the local zonal wind trend and the shifts in the probability density between wave regimes, via the changes in local $C_{gx}$, may provide an insight into this issue. Future studies could further evaluate how these local $C_{gx}$ trends are driven by Arctic and tropical warming, as well as sea surface temperature anomalies, and ultimately alter regional surface weather and climate.

## Research Funding

J.L. and Z.L. appreciate the support from the National Natural Science Foundation of China (Grant No. 42175070). Y.L. and Z.L. were also supported by the National Natural Science Foundation of China (Grant No. 42288101). T.W. was supported by NERC grant NE/C001288/1. This study was supported by the National Key Scientific and Technological Infrastructure project "Earth System Numerical Simulation Facility" (EarthLab).

## Data Availability Statement

The ERA5 data were downloaded from https://apps.ecmwf.int/data-catalogues/era5/?class=ea. The CESM2-LE is available at https://www.cesm.ucar.edu/community-projects/lens2.

## Author contributions

Z.L. developed the methodology, performed the dignostics and visualization. J.L. initialized the ideas on group velocity trends and shift of Rossby wave regimes and conceptualize the study. T.W. proposed the idea of using large ensemble experiments. Z.L., J.L. drafted the original manuscript. J.L. and Y.L. supervised the research. All of the authors participated in the discussion, reviewed and edited the manuscript.

Supplementary Information for

# Shifts in Rossby wave regimes via the change of group velocity with zonal jets

**Zhixiang Li[1,2], Jianhua Lu[1], Yimin Liu[2], and Tim Woollings[3]**

[1]School of Atmospheric Sciences, Sun Yat-sen University and Southern Marine Science and Engineering Guangdong Laboratory (Zhuhai), Zhuhai, China

[2]Key Laboratory of Earth System Numerical Modeling and Application, Institute of Atmospheric Physics, Chinese Academy of Sciences, Beijing, China

[3]Atmospheric, Oceanic and Planetary Physics, University of Oxford, Oxford, United Kingdom

Corresponding author: J. H. Lu (lvjianhua@mail.sysu.edu.cn), Y. M. Liu (lym@lasg.iap.ac.cn)

The Supplementary Information includes:

Supplementary Figures 1-12

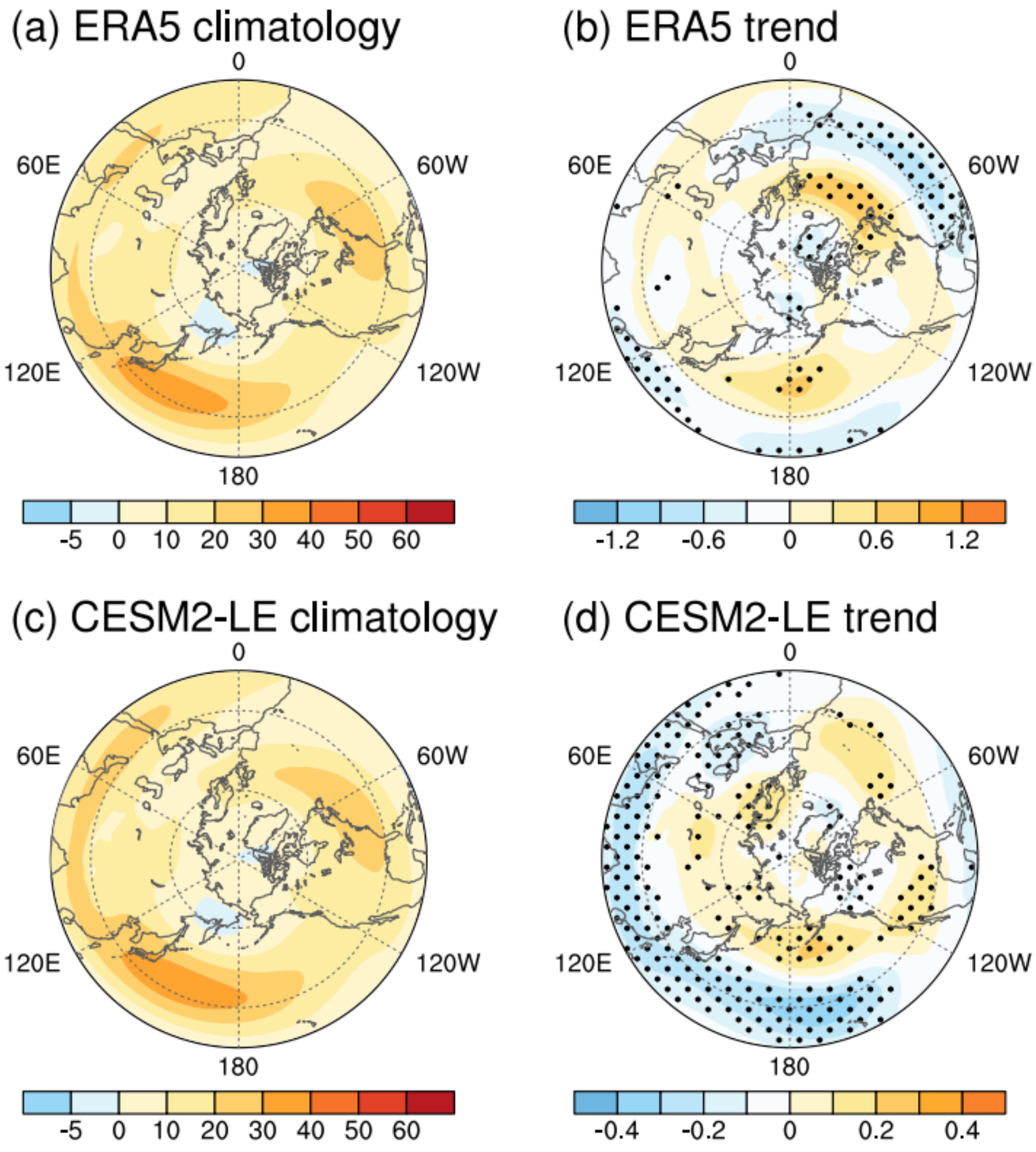

**Figure S1.** Climatology and trends of zonal wind at 500 hPa (U500) during boreal winter from 1954 to 2023. As in Fig. 1 but for results at 500 hPa.

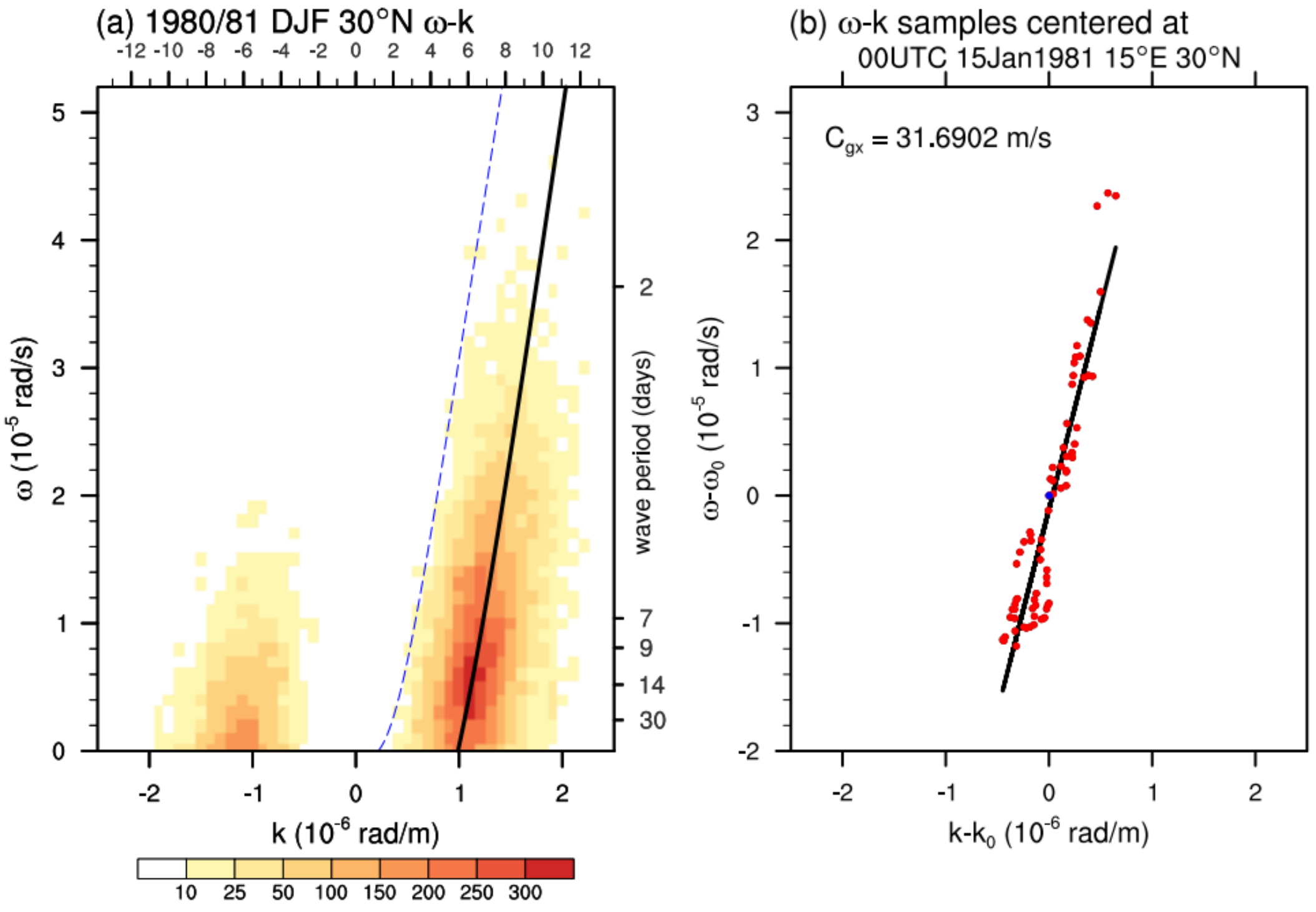

**Figure S2.** (a) The local angular frequency–local zonal angular wavenumber ($\omega$–$k$) distribution during 1980/81 DJF at 30°N, color shading indicates the numbers of $\omega$–$k$ pairs within a certain bin ($1 \times 10^{-6}$ for $\omega$ and $1 \times 10^{-7}$ for $k$). The thick black line indicates the quasi-geostrophic Rossby wave dispersion relation, while the thin dashed blue line is barotropic Rossby wave dispersion relation. (b) A typical example for calculating $\partial\omega/\partial k$ located at 15°E, 30°N at 00UTC 15 January 1981. All $\omega$–$k$ pairs within a spatial range of 3 gridpoints $\times$ 3 gridpoints and within a temporal range of $\pm$24 hours are shown. The blue dot represents the $\omega_0$–$k_0$, while the red dots indicate the $\omega$–$k$ pairs satisfy $|(k-k_0)/k_0| < 0.5$. The black line denotes the linear regression line and the significant linear regression coefficient is 31.6902 m/s.

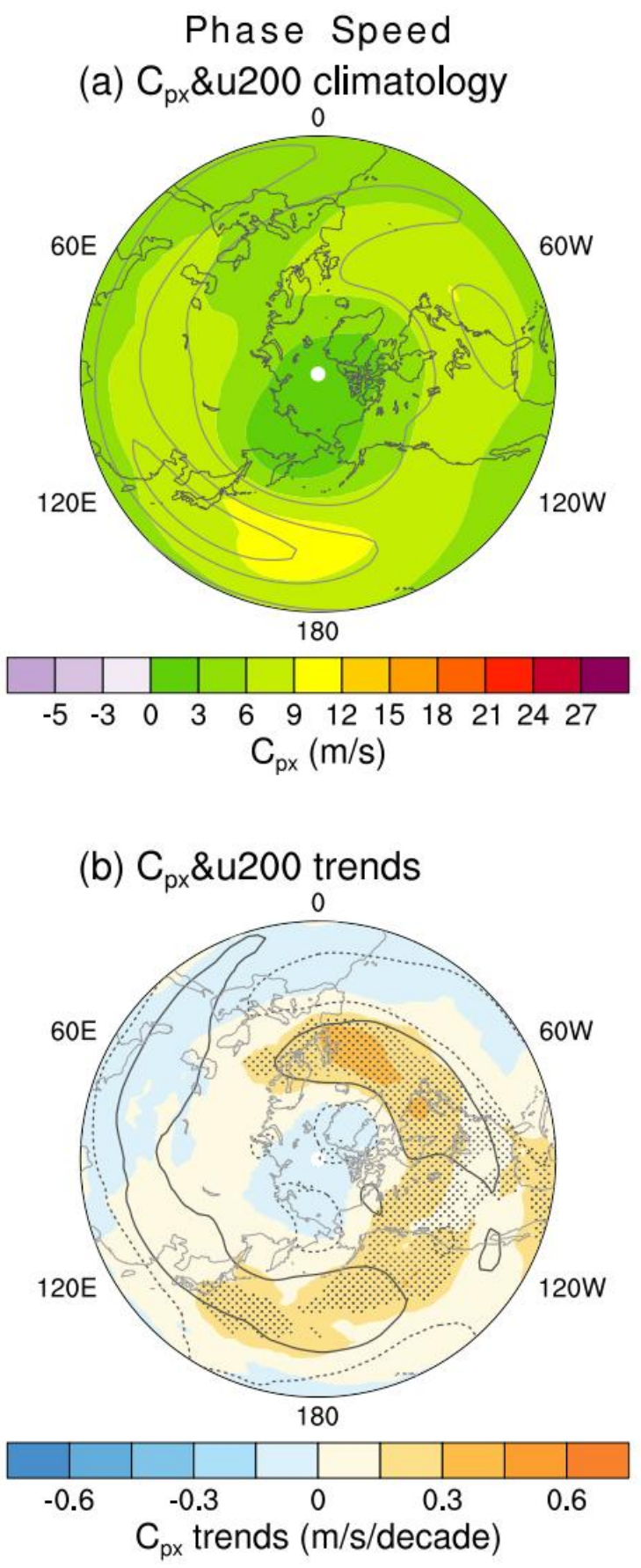

**Figure S3.** Observed climatology (a; units: m s$^{-1}$) and long-term trends (b; units: m s$^{-1}$ (10 yr)$^{-1}$) of $C_{px}$ during DJF over 1954–2023 at 200 hPa. Black contours indicate the corresponding climatology (a, only 20, 40, 60 m s$^{-1}$) and trends (b, only $\pm$0.5 m s$^{-1}$ (10 yr)$^{-1}$; negative contours dashed) of $u$. Stippling indicates the trends in $C_{px}$ are statistically significant above the 95% confidence level. The color bar is the same as that of group velocity in Figure 3.

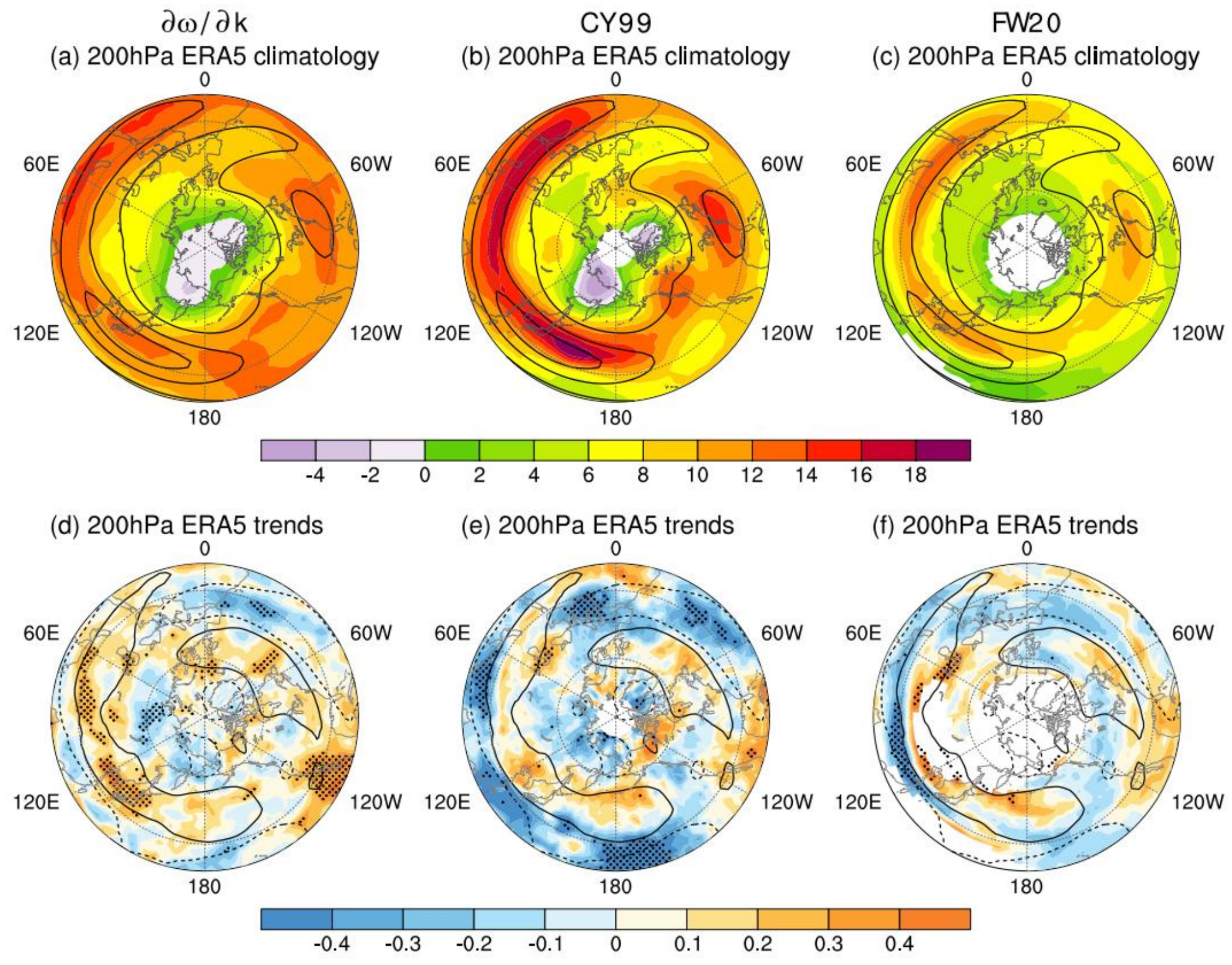

**Figure S4.** Observed climatology and long-term trends of extratropical Rossby wave dispersion during DJF over 1954–2023 at 200 hPa. Climatology (a–c; units: m $s^{-1}$) and trend (d–f; units: m $s^{-1}$ (10 yr)$^{-1}$) of DJF mean Rossby wave dispersion derived from our new method (left panel), CY99 method (middle panel) and FW20 method (right panel), respectively. Black contours indicate the corresponding climatology (a–c, only 20, 40, 60 m $s^{-1}$) and trends (d–f, only $\pm$0.5 m $s^{-1}$ (10 yr)$^{-1}$; negative contours dashed) of DJF $u$. Stippling indicates the trends in dispersion are statistically significant above the 95% confidence level. The boundary circle is 20°N.

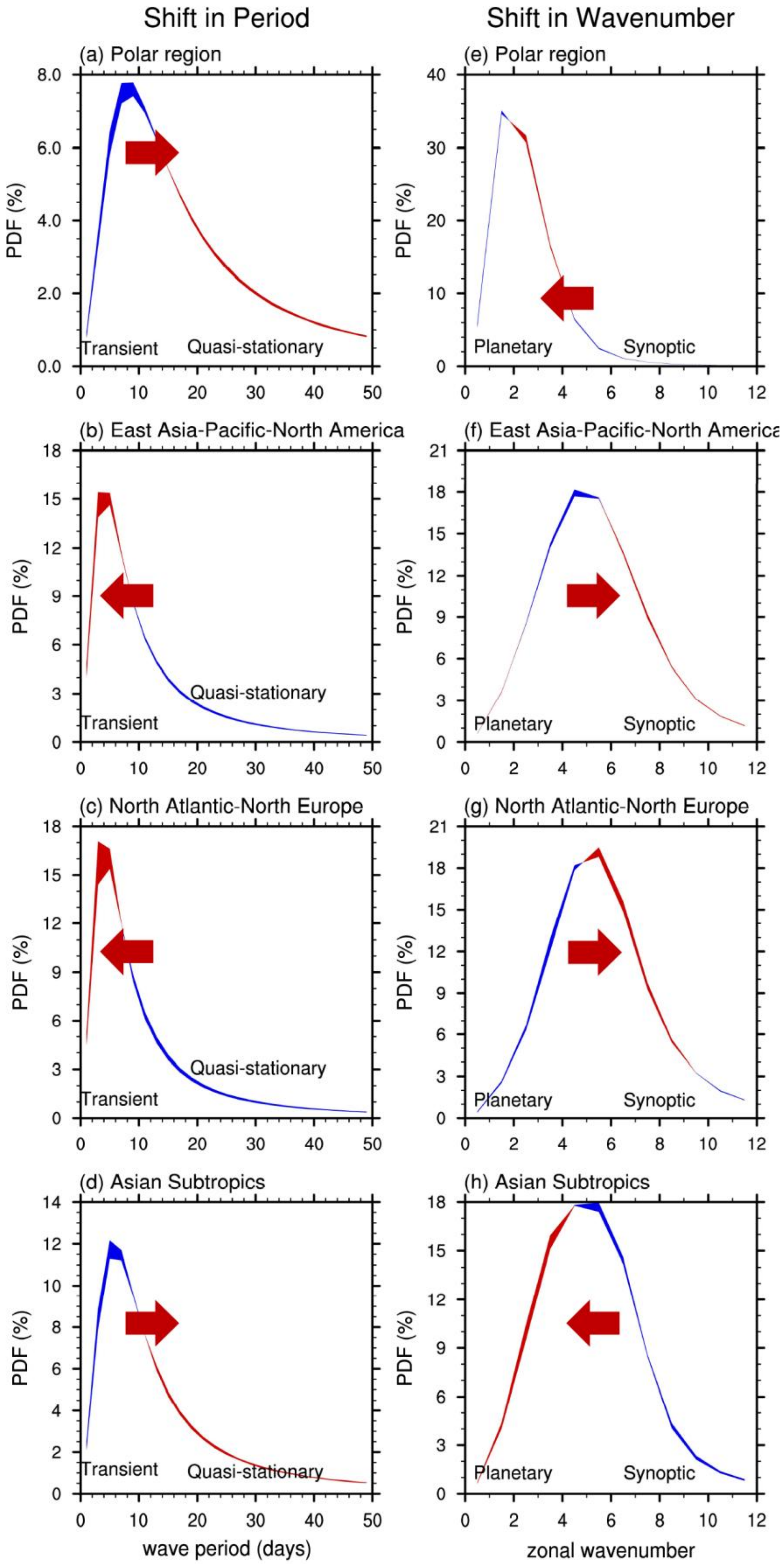

**Figure S5.** Observed shifts in the spatiotemporal-scale regimes of extratropical Rossby waves during DJF over 1954–2023 at 200 hPa. Shading are the differences in the PDF of wave period (a–d) and zonal wavenumber (e–h) between the last 30 years (1993/94–2022/23) and first 30 years (1954/55–1983/84) for the polar region (a, e; 70°–325°E, 62.5°–85°N), the East Asia-North Pacific-northwest North America sector (b, f; 110°–250°E, 35°–57.5°N), the North Atlantic (c, g; 280°–357.5°E, 37.5°–62.5°N), and the South Asian subtropics (d, h; 60°–135°E, 22.5°–31°N). Red shading indicates increases in PDF and blue shading indicates decreases in PDF.

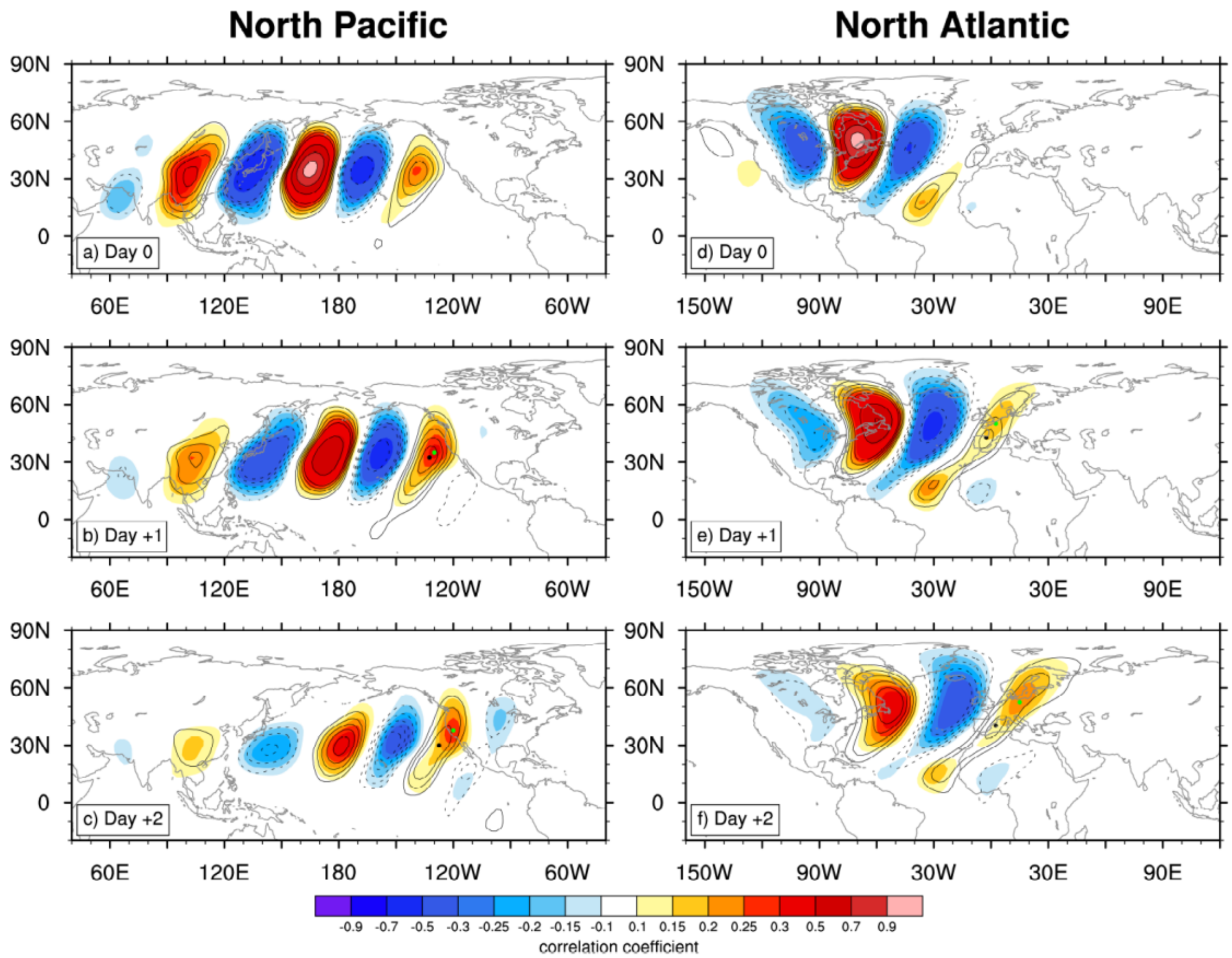

**Figure S6.** Observed changes in the evolutionary characteristics of Rossby waves over the Pacific–North America sector (left panels) and the North Atlantic–Europe sector (right panels), based on time-lagged one-point correlation from day 0 to day +2. The shading represents the averaged correlation coefficients for the last 30 winters (1993/94–2022/23), while contours show the averaged correlation coefficients for the first 30 winters (1954/55–1983/84; contour interval identical to the shading; negative contours dashed). Results are calculated using the synoptic-scale (wavenumbers 1–15) *v* during DJF, with base points at 190°E, 30°N (290°E, 50°N). Because the Pacific–North America sector is predominantly controlled by synoptic-scale waves (Fig. 5b), the signals are more clearly revealed when using only synoptic-scale *v*.

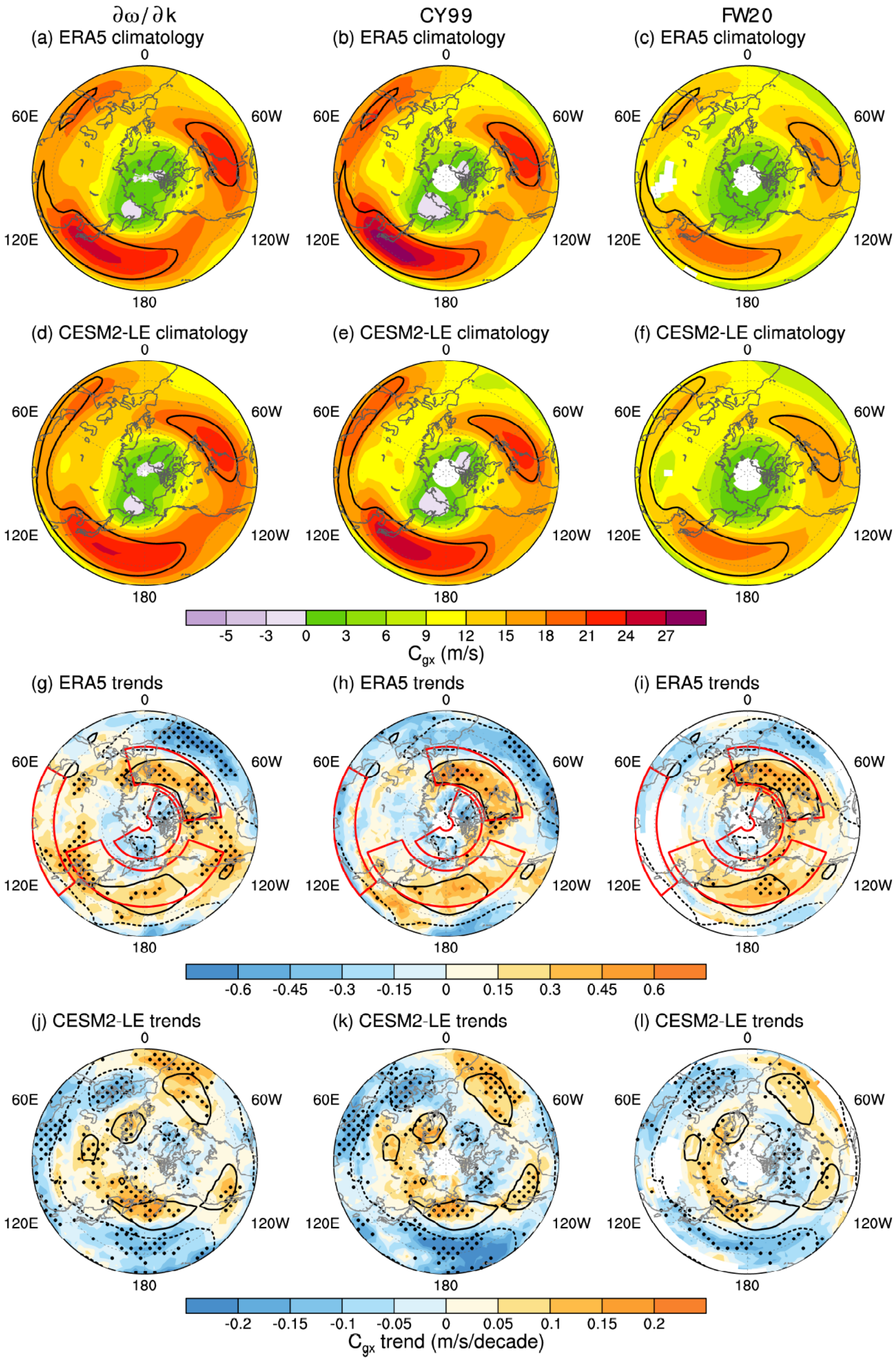

**Figure S7.** Observed and modeled climatology and long-term trends in $C_{gx}$ and $u$ at 500 hPa. As in Fig. 3 but for results at 500 hPa.

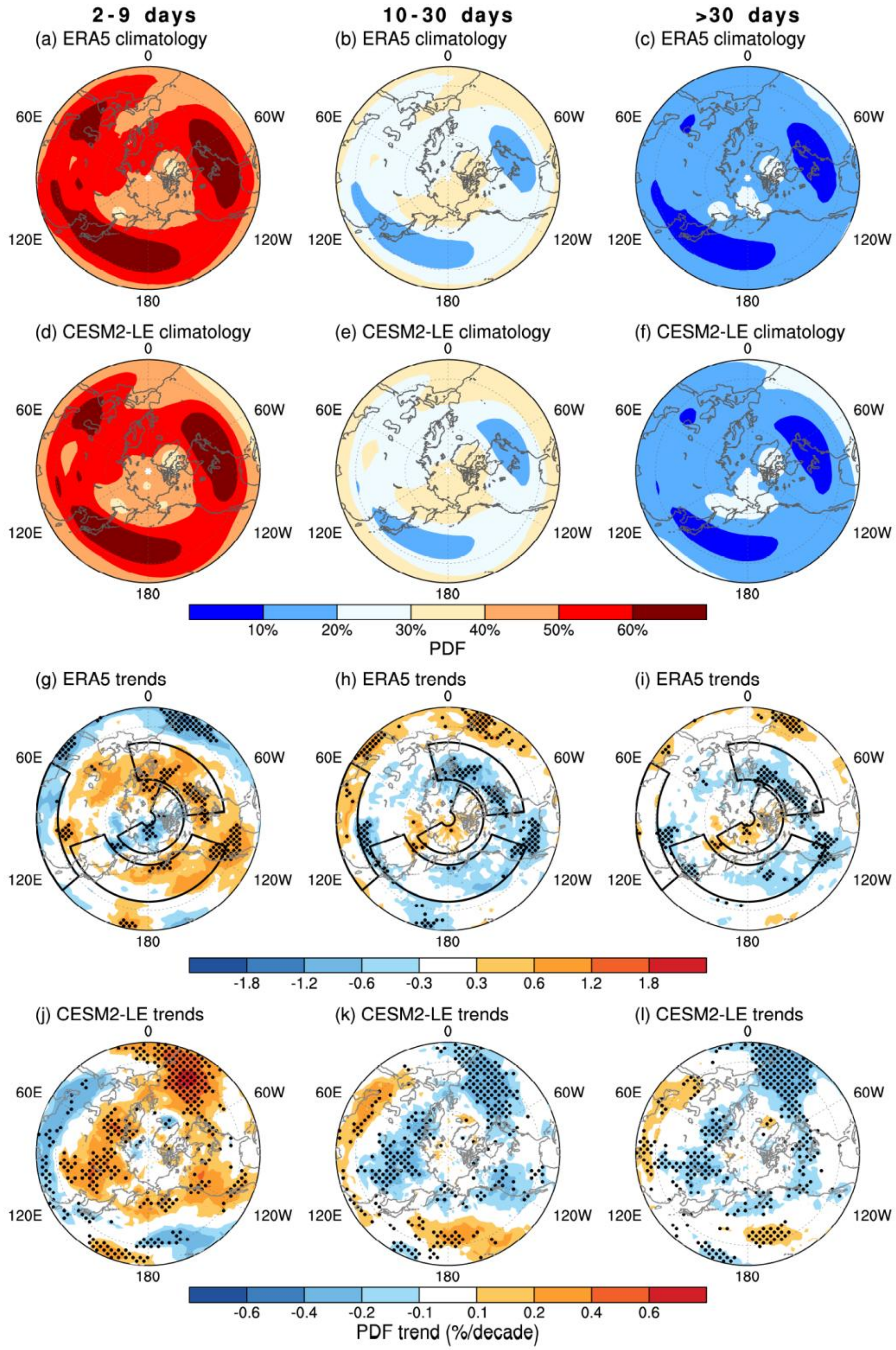

**Figure S8.** Observed and modeled climatology and long-term trends in the PDF of wave periods at 500 hPa. As in Fig. 4 but for results at 500 hPa.

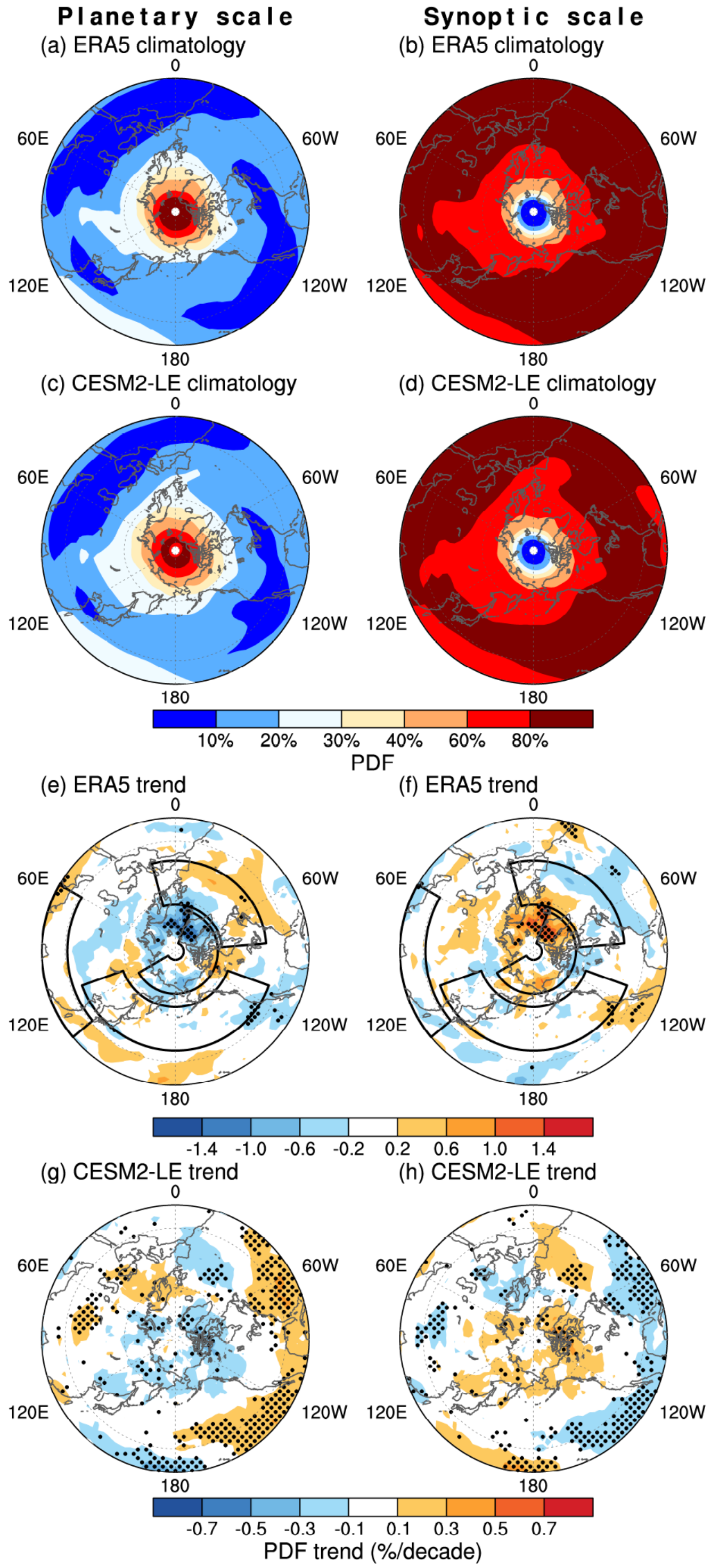

**Figure S9.** Observed and modeled climatology and long-term trends in the PDF of wavenumbers at 500 hPa. As in Fig. 5 but for results at 500 hPa.

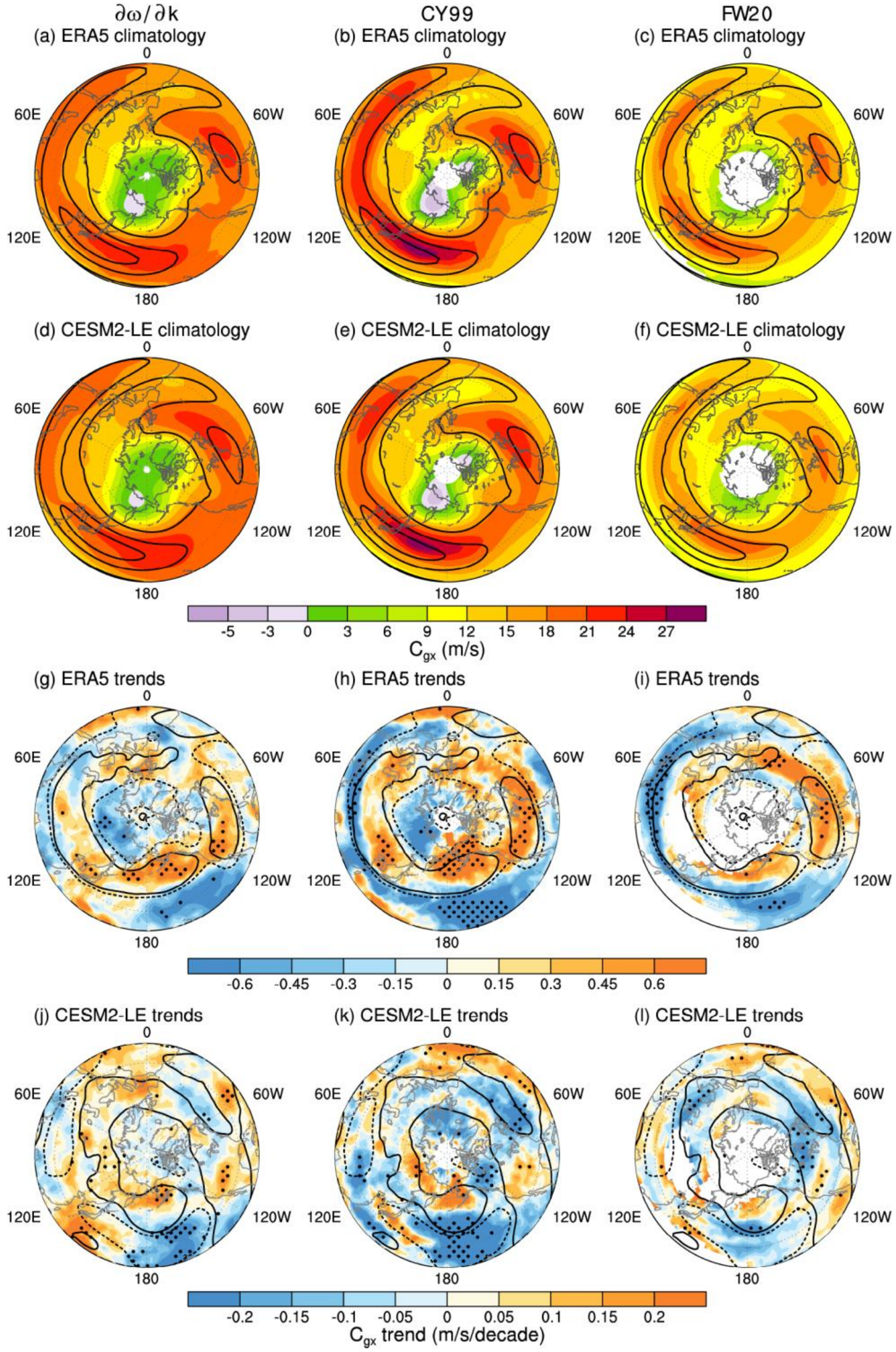

**Figure S10.** Observed and modeled climatology and trends in $C_{gx}$ and $u$ during the satellite era (1979–2023). As in Fig. 3 but for the period 1979–2023.

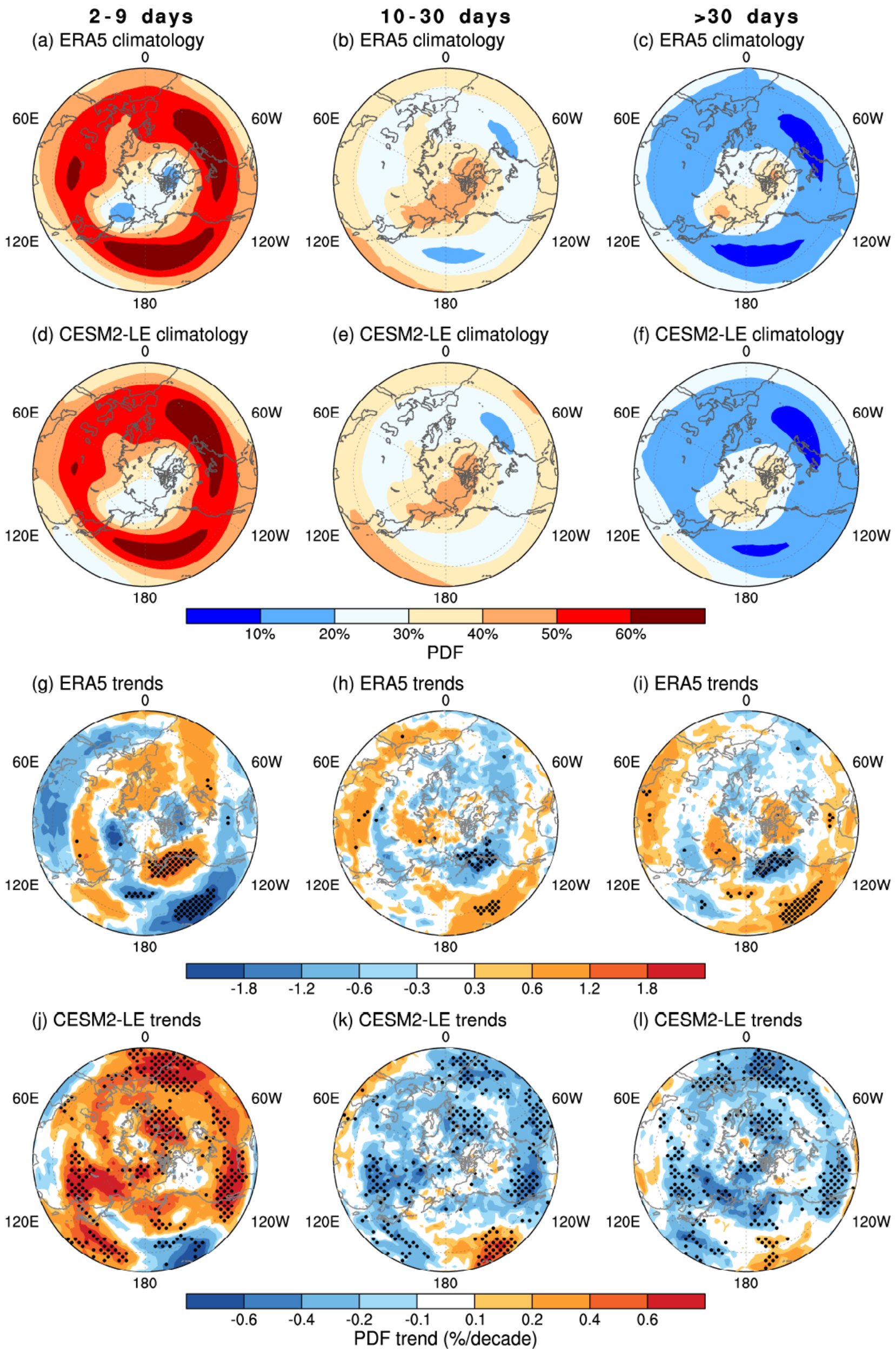

**Figure S11.** Observed and modeled climatology and trends in the PDF of wave periods during the satellite era (1979–2023). As in Fig. 4 but for the period 1979–2023.

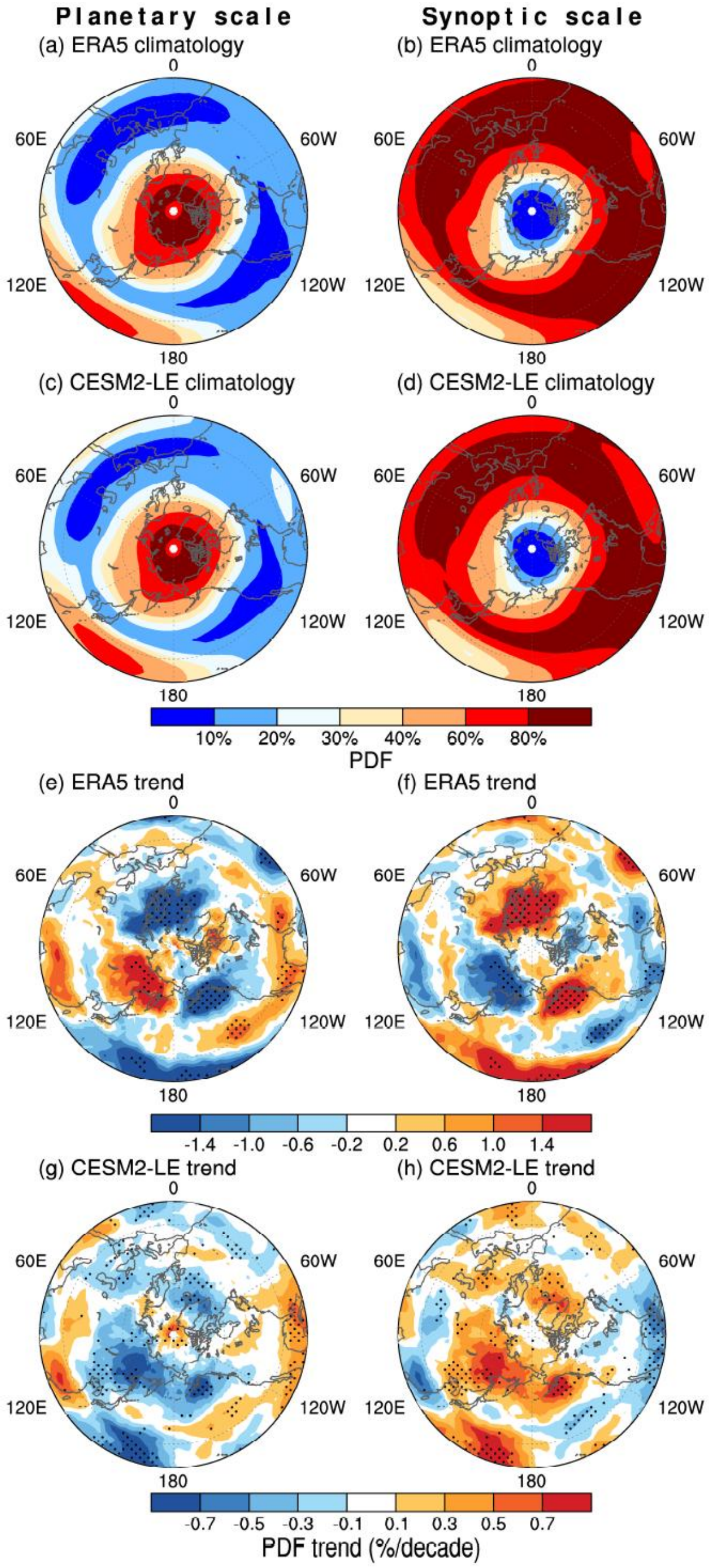

**Figure S12.** Observed and modeled climatology and trends in the PDF of wavenumbers during the satellite era (1979–2023). As in Fig. 5 but for the period 1979–2023.